\documentclass[letterpaper, 12pt]{article}

\usepackage[english]{babel}
\usepackage[utf8x]{inputenc}
\usepackage[T1]{fontenc}
\usepackage{times}

\usepackage[a4paper,top=2cm,bottom=2cm,left=2cm,right=2cm,marginparwidth=1.75cm]{geometry}
\usepackage{setspace}   

\usepackage{amsmath}
\usepackage{amssymb}
\usepackage{multirow}
\usepackage{graphicx}
\usepackage{natbib}
\usepackage{wasysym} 
\usepackage{subcaption}
\PassOptionsToPackage{hyphens}{url}\usepackage[colorlinks=true, allcolors=red]{hyperref}
\hypersetup{breaklinks=true}

\usepackage{lscape}
\usepackage{listings}
\newcommand{\papertitle}{{COVID-19 and the gig economy in Poland}}  

\title{\papertitle}
\author{Beręsewicz Maciej\footnote{Corresponding author: \href{mailto:maciej.beresewicz@ue.poznan.pl}{maciej.beresewicz@ue.poznan.pl}. Poznań University of Economics and Business, Al. Niepodległości 10,
61-875 Poznań, Poland. Statistical Office in Poznań, Wojska Polskiego 27/29,
60-624 Poznań, Poland  } \\
\textit{Poznań University of Economics and Business, Poland} \\
\textit{Statistical Office in Poznań, Poland} \\

Nikulin Dagmara\footnote{\href{mailto:dnikulin@zie.pg.gda.pl}{dnikulin@zie.pg.gda.pl}. Gdańsk University of Technology, Faculty of Management and Economics, Narutowicza 11/12, 80-233 Gdańsk, Poland } \\
\textit{Gdańsk University of Technology, Poland}  \\

}

\date{}

\begin{document}


\maketitle

\doublespacing

\begin{abstract}
We use a dataset covering nearly the entire target population based on passively collected data from smartphones to measure the impact of the first COVID-19 wave on the gig economy in Poland. In~particular, we focus on transportation (Uber, Bolt) and delivery (Wolt, Takeaway, Glover, DeliGoo) apps, which make it possible to distinguish between the demand and supply part of this market. Based on Bayesian structural time-series models, we estimate the causal impact of the first COVID-19 wave on the number of active drivers and couriers. We show a~significant relative increase for Wolt and Glover (15\% and 24\%) and a~slight relative decrease for Uber and Bolt (-3\% and -7\%) in comparison to a~counterfactual control. The change for Uber and Bolt can be partially explained by the prospect of a new law (\textit{the so-called Uber Lex}), which was already announced in 2019 and is intended to regulate the work of platform drivers.
\end{abstract}


\vspace{0.5cm}


\vspace{0.5cm}

\noindent \textbf{Number of words}: 1999


\clearpage





\clearpage

\section{Introduction}



The COVID-19 pandemic has significantly influenced many areas of life, including the labor market situation. Periods of domestic isolation and lockdown-induced job losses may have contributed to the development of the platform economy, which is an increasingly large part of the labor market. The platform economy (hereafter also gig economy) can be defined as ``non-standard work facilitated by online platforms, which use digital technologies to 'mediate' between individual suppliers (platform workers) and buyers of labour'' \citep[p. 98]{Hauben2020}. However, existing studies report that COVID-19 has had a varying impact on the extent of the gig economy. At the same time, many empirical studies are based on online surveys and do not differentiate between specific types of gig work, such as, crowd workers, who perform tasks online, highly skilled gig workers, such as architects or software engineers, and low skilled gig workers, like drivers and deliverers \citep{spurk2020flexible}.  Importantly, if the separate sectors of the gig economy are taken into account, the impact of COVID-19 may prove to be differential \citep{cao2020impact}. 


In~this article we focus on gig jobs performed by drivers and couriers, so we narrow down the scope of the gig economy to the transportation and delivery sectors. In~this field recent evidence suggests an increase in tasks posted and filled within the platform economy during the pandemic, as~a~result of many people working remotely and the closure of restaurants. In particular, if one considers the sector of food delivery, it is possible to observe an increasing trend, see among others \citet{batool2020covid, raj2021covid}. As far as transportation services are concerned, the pandemic may have had a negative impact, which is documented by \citet{batool2020covid} using Google Trends for Uber and Lyft. Similarly, due to the social distancing measures and lockdowns earnings of drivers and couriers have decreased.



As regards Poland, \citet{polkowska2021platform} analysed the impact of the first wave of COVID-19 on the work of Glovo couriers. Based on 20 semi-structured interviews and 1300 posts in the Internet forum, she concludes that during the pandemic, working as a~courier is perceived as a good occupation, which can offer opportunities for those who lost their job during the lockdown. Moreover, the popularity of such activities is increasing following a growing number of orders. Similarly, \citet{muszynski2021} find an increasing interest in food delivery platforms (Glovo, Takeaway and Stava), which reflects the surge in demand for such services.  The main limitations of the above-mentioned studies are their qualitative nature and the fact they do not attempt to estimate the effect of COVID-19 on this sector.

In~this article we take a~different approach using a dataset that covers over 21 million smartphone users in Poland (Poles and foreigners) to study the impact of COVID-19\footnote{In~2020 the population of people aged 18 and older in Poland consisted of about 31 million Polish citizens and about 1.5 million foreigners. About 97\% of them claimed to have a~phone and 80\% of them reported using a~smartphone. This suggests that in 2020 there were about 24 million smartphone users in Poland, so the data we use covers over 85\% this population.}. Our main research question is: \textit{What is the size of the effect of the first COVID-19 wave on the gig economy, in particular on the transportation and food delivery services?}. Our results contribute significantly to the existing knowledge by providing empirical quantitative evidence about the impact of the COVID-19 pandemic on the gig economy. Unlike previous studies, which are mostly based on surveys or interviews, our study makes it possible to track changes on the gig labour market more precisely. The article has the following structure. In~Section \ref{sec-dat} we describe trends in the number of active users of 6~apps used in the study. In~section \ref{sec-meth} we briefly describe a Bayesian structural time-series model proposed by \citet{brodersen2015inferring}, which we use to estimate the causal effect of the first COVID-19 wave. Section \ref{sec-results} presents the results and is followed by conclusions.

\section{Data}\label{sec-dat}

To distinguish between the demand and the supply part of the gig economy we focused only on apps specially created for workers i.e. drivers and couriers. We identified the following mobile apps for couriers: \textit{Takeaway}, \textit{Glover}, \textit{Wolt} and \textit{DeliGoo} and for drivers: \textit{Bolt}, \textit{Uber}. However, in the latter case, it is not possible to distinguish between drivers and couriers, since the same app is used to provide two kinds of services: Uber Driver and Uber Eats. A detailed description of the data can be found in \citet{beresewicz2021}.

\begin{figure}[ht!]
    \centering
    \includegraphics[width=\textwidth]{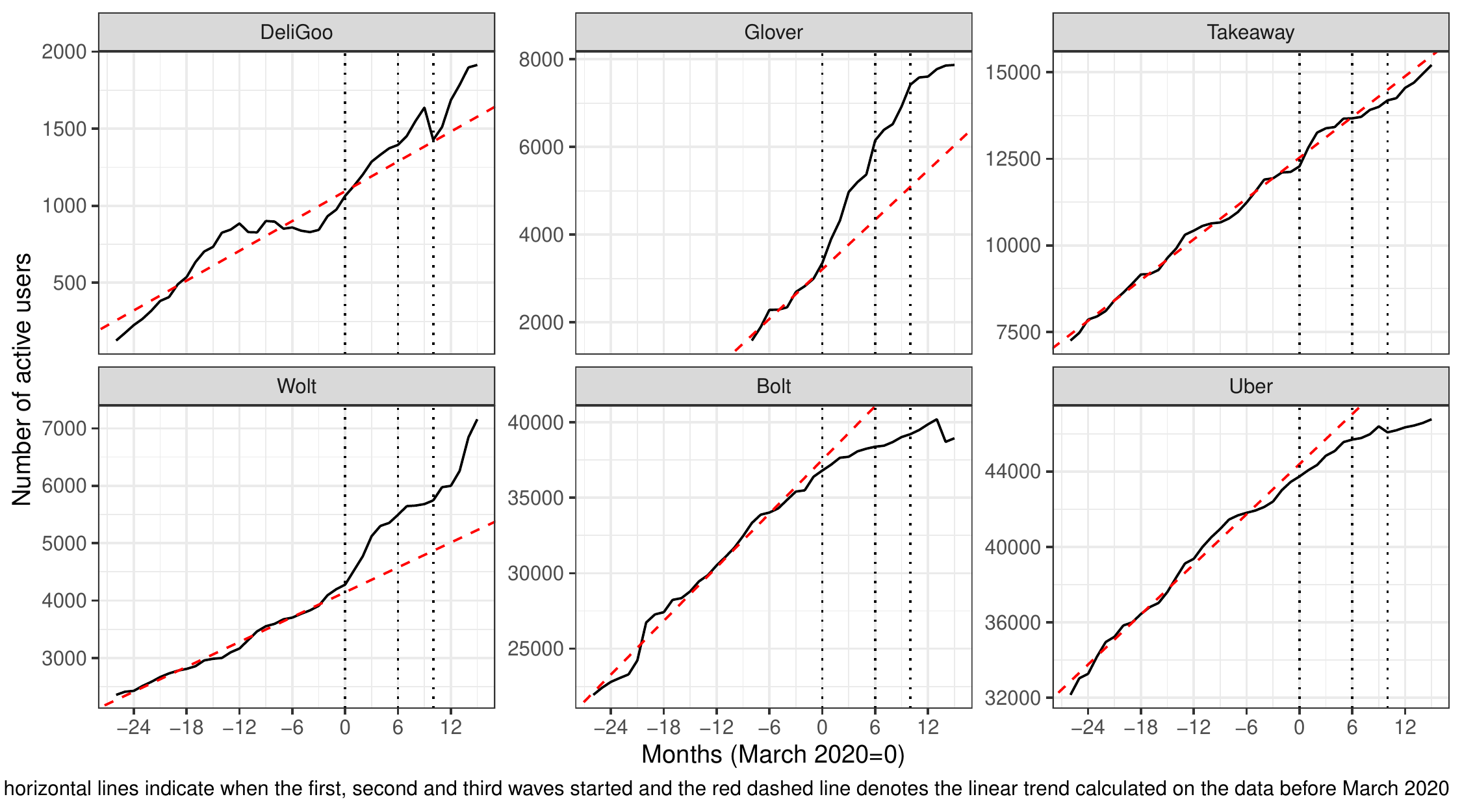}
    \caption{The monthly number of active users in Poland for selected apps between January 2018 and June 2021}
    \label{fig-general}
\end{figure}

Figure \ref{fig-general} presents the number of active users of the 6 apps between January 2018 and June 2021. The number of months before and after the first wave of COVID-19, which started in March 2020, are presented on the X axis. The number of COVID-19 cases in Poland can be found in Figure \ref{fig-covid-19} included in the Supplementary Materials. The dotted horizontal lines indicate when the first, second and third waves started and the red dashed line denotes the linear trend calculated on the data before March 2020. 

For Glover and Wolt we observe a~sharp increase after the beginning of the first wave. For Takeaway there is a~slight increase but it is not significantly different from the pre-COVID-19 linear trend. The number of  active users of DeliGoo varied over the reference period and we can observe an increase starting at the beginning of 2020, which may be connected with the company's advertising campaign. For Bolt and Uber the trend is different: there is  a~decrease, which is larger for Bolt. This may be due to the character of the Uber app, which includes drivers and couriers. The decline for Bolt and Uber seems to start at the end of 2019 and the beginning of 2020, which may be the result of the prospect of a new law intended to regulate the work of platform drivers. 

\section{Methods}\label{sec-meth}

In~order to estimate the causal effect of the first COVID-19 wave we use Bayesian structural time-series models proposed by \citet{brodersen2015inferring}. This method involves a~diffusion-regression state-space model to predict the counterfactual response in a~synthetic control that would have occurred had no intervention taken place. The underlying model consists of three parts: the local linear trend, seasonality and contemporaneous covariates with static or dynamic coefficients. We did not assume the seasonal component as our time series is too short (2.5 years) and its deviations from the linear trend are negligible.

The estimation of a causal effect requires covariate(s) that will be used for counterfactual analysis. As COVID-19 impacted almost all monthly macroeconomic indicators, we decided to use the number of active Takeaway couriers as a control. The motivation for this choice is twofold: first, it does not significantly differ from the pre-COVID-19 trend and despite a~slight increase after the first wave, it quickly returned to the pre-COVID trend; second, Takeaway had the same popularity during the first wave\footnote{See Google Trends in section \ref{appen-google-trends} in the Supplementary Materials.} and couriers may in fact be restaurant employees rather than freelance workers. Table \ref{tab-demo-users} in the Supplementary Materials provides background information about gig workers, which indicates that there is no significant difference between those who worked before and after the first wave.

\section{Results}\label{sec-results}

Table \ref{tab-estims} presents point and relative estimates of the causal effect of the first COVID-19 wave calculated at the end of August 2020, just before the second wave. According to our approach, a significant, positive impact was estimated for users of the Glover and Wolt apps, which are dedicated to delivery services. The point estimate is 1,281 for Glover and 814 for Wolt, which in relative terms represent a 24 and 15 percent increase, respectively.  In addition to ready-made dishes, Glover couriers also deliver groceries. From the beginning, the Glovo company cooperated with the largest chain of supermarkets in Poland (Biedronka) and fast foods (e.g. McDonald's). Both apps are particularly popular in cities, where the demand is the highest. In the case of DeliGoo, the number of couriers did not change significantly, which may be due to its low popularity.

\begin{table}[ht!]
\centering
\begin{tabular}{lrrrr}
  \hline
  & \multicolumn{2}{c}{Absolute} & \multicolumn{2}{c}{Relative [in \%]} \\
  App & Estimate & 95\% CI & Estimate & 95\% CI \\ 
  \hline
  DeliGoo &  88 & (-173; 339) & 6.4 & (-12.6; 24.7) \\ 
  Glover & 1,281 & (561; 2,031) & 23.8 & (10.4; 37.8) \\ 
  Wolt & 814 & (322; 1,288) & 15.2 & (6.0; 24.1) \\ 
  Bolt & -2,595 & (-6,576; 1,260) & -6.8 & (-17.2; 3.3) \\
  Uber &  -1,324 & (-4,245; 1,621) & -2.9 & (-9.3; 3.6) \\
   \hline
\end{tabular}
\caption{Absolute and relative point estimates of the causal effect of the first COVID-19 wave in Poland in August 2020}
\label{tab-estims}
\end{table}

As expected, the number of active users for Bolt and Uber dropped by 2.6 thousand and 1.3 thousand, which, in relative terms, represents 6.8\% and 2.9\%, respectively. For both apps the credible interval is quite large but asymmetric relative to zero, which suggests a high probability of the negative impact of COVID-19. The difference in the decline for Bolt and Uber may be due to fact that the Uber app can be used by drivers and couriers. Furthermore, the negative trend for Uber and Bolt began before the pandemic following the announcement at the end of 2019 of on a new law (the so-called \textit{Uber Lex}), which was intended to regulate the work of platform drivers. The Uber Lex came into effect in 2020 with a~transition period until the end of March, which was then extended to the end of September and, later, to the end of December owing to the pandemic.

\section{Conclusions}

The purpose of this article was to analyse the impact of the first COVID-19 wave on the gig economy, in particular the transportation and food delivery sectors. Our results show considerable differences among gig workers engaged in transportation and food delivery sectors. In other words, the particular sub-sectors of the gig economy have been affected in different ways by the coronavirus crisis. Social distancing and lockdown measures introduced in March 2020 caused a halt in the development of the transport-related gig sectors of the economy in Poland. This means that the negative impact of the pandemic on transportation gig workers in Poland is similar to the effects observed in other countries \citep{batool2020covid, katta2020dis}. At the same time, the extraordinary situation where restaurants and bars were closed and many people had to self-isolate in homes has triggered a growing demand for food delivery services, which led to a sharp rise in the number of couriers, as confirmed by recent evidence provided by \citet{batool2020covid} and \citet{raj2021covid}.

Our study has a number of potential policy implications. First, we contribute to the existing knowledge by providing evidence about the varying impact of the pandemic on separate sectors of the gig economy. It supports the hypothesis that the gig economy is very heterogeneous and cannot be analysed as a whole. Moreover, given the increase in the popularity of services offered by platform economy, the crisis caused by the pandemic highlighted the uncertain and unstable situation of gig workers, who cannot rely on a steady source of income. As incomes of gig workers are unstable and workers are often not covered by the labor protections offered to employees, the pandemic crisis has been exacerbated as a result of these precarious employment types. However, it is worth mentioning that the social reaction including pressure from regulators, driver advocates, and the media has helped to improve the social protection of Uber drivers in terms of safety and the mitigation of health risks \citet{katta2020dis}. 





\section*{Acknowledgements}

The study was conducted as part of the research project \textit{Economics in the face of the New Economy}, financed under the Regional Initiative for Excellence programme of the Minister of Science and Higher Education of Poland, years 2019-2022, grant no. 004/RID/2018/19, financing 3,000,000 PLN (for Maciej Beręsewicz).  Data were obtained from the Selectivv company (\url{https://selectivv.com/en/}). The views expressed in this article are those of the authors and do not necessarily reflect the official policy or position of Statistics Poland or the Statistical Office in Poznań. The authors would like to thank Jakub Sawulski for his valuable comments that led to the creation of this article; We also thank Marcin Augustyniak for explaining the data and Robert Pater for his comments on the early version. All data and codes are available at \url{https://github.com/DepartmentOfStatisticsPUE/rid-gig-economy-covid19}.


\bibliographystyle{apalike}
\bibliography{bibliography}

\appendix

\clearpage

\begin{center}
    \Large Supplementary materials to the article\\
    \textit{\papertitle}
\end{center}

\section{COVID-19 in Poland}

\begin{figure}[ht!]
    \centering
    \includegraphics[width=\textwidth]{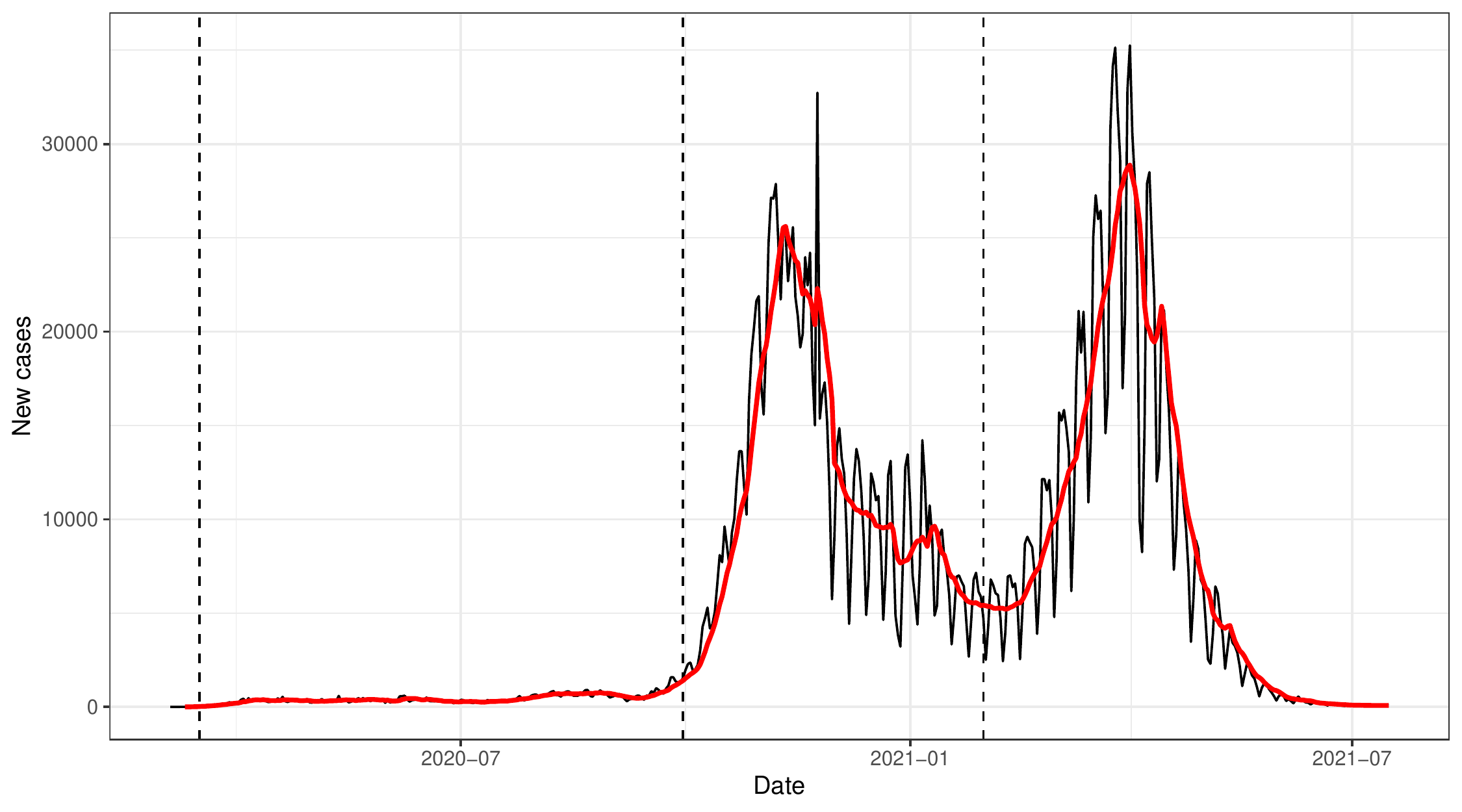}
    \caption{The number of COVID-19 cases in Poland. The dashed lines denote the beginning of the first (2020-03-16), second (2020-09-30) and third (2021-01-31) wave. The first lockdown was introduced  on 2020-03-18 }
    \label{fig-covid-19}
\end{figure}

Important dates regarding the first lockdown in Poland:

\begin{itemize}
    \item 12 March -- closure of all schools until March 25, which was then extended until April 10 - classes were held remotely, universities switched to distance learning, the activity of cultural institutions, i.e. philharmonics, operas, theaters, museums, cinemas, was suspended.
    \item 14 March -- the functioning of shopping centers and galleries was limited. Only grocery stores, pharmacies, drugstores, and laundries remained open.
    \item 15 March --  borders closed for air and rail traffic.
    \item 25 March  -- limitation of assemblies to max. 2 people (except for families, religious gatherings and workplaces), only essential travel was permitted, people were allowed to go outside only in essential situations, e.g. shopping, to buy medicines, visit a doctor, walk the dog, physical activity in the fresh air; bars, restaurants, pubs, casinos, cinemas, theaters and other entertainment venues remained closed.
    \item 31 March -- children under the age of 18 were not allowed to leave homes without a guardian; parks, boulevards and beaches were closed, as well as hairdressing, beauty, tattoo and piercing salons; hotels could only operate if guests were quarantined or were staying on business, such as medics or construction workers; the obligation to keep a distance of 2 meters from each other was introduced, except for caretakers of children under the age of 13 and disabled persons.
    \item 1 April --  the number of customers in~shops and service outlets could not exceed three times the number of cash registers (double counters in the case of post offices).
\end{itemize}

\clearpage

\section{Google trends}\label{appen-google-trends}

\begin{figure}[ht!]
    \centering
    \includegraphics[width=\textwidth]{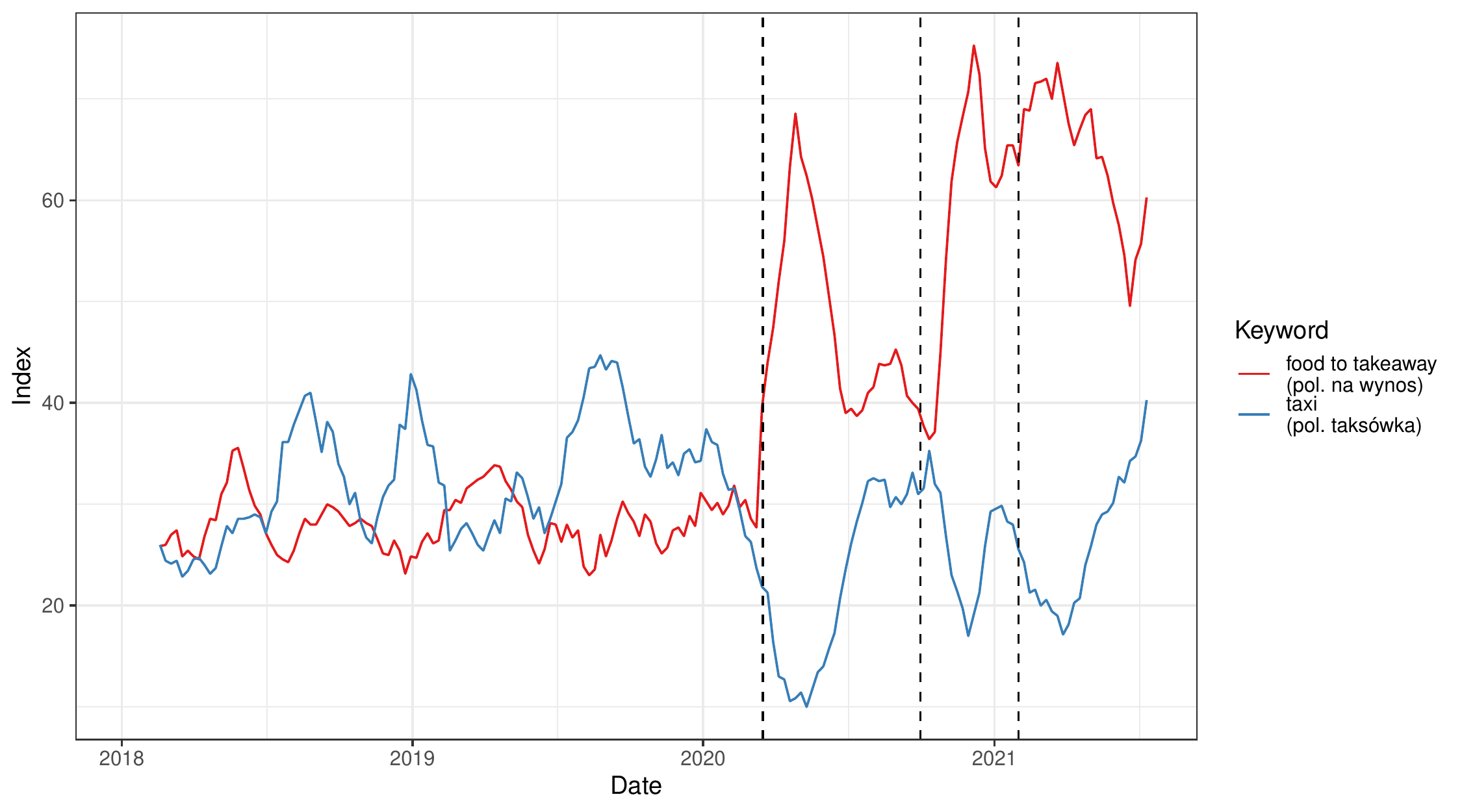}
    \caption{Popularity of two keywords: food to takaway and taxi in Google Trends. Each line denotes a 7-day moving average and dashed horizontal lines denote the beginning of a given COVID-19 wave}
    \label{gtrends-1}
\end{figure}

\begin{figure}[ht!]
    \centering
    \includegraphics[width=\textwidth]{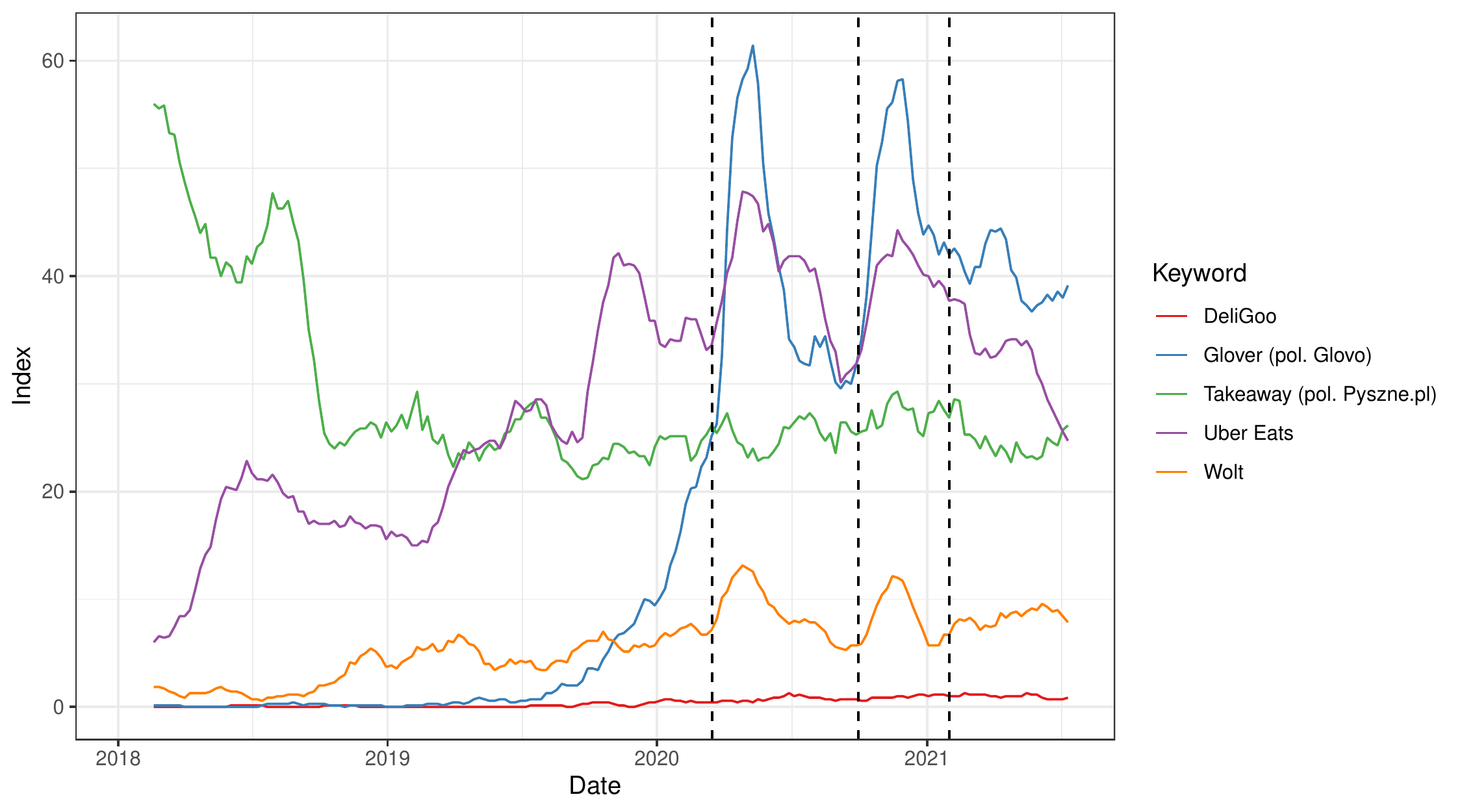}
    \caption{Popularity of five keywords regarding delivery services in Google Trends. Each line denotes a 7-day moving average and dashed horizontal lines denote the beginning of a given COVID-19 wave. Note that in Polish the search word "Wolt" may also refer to a~unit for electric potential.}
    \label{gtrends-2}
\end{figure}

\begin{figure}
    \centering
    \includegraphics[width=\textwidth]{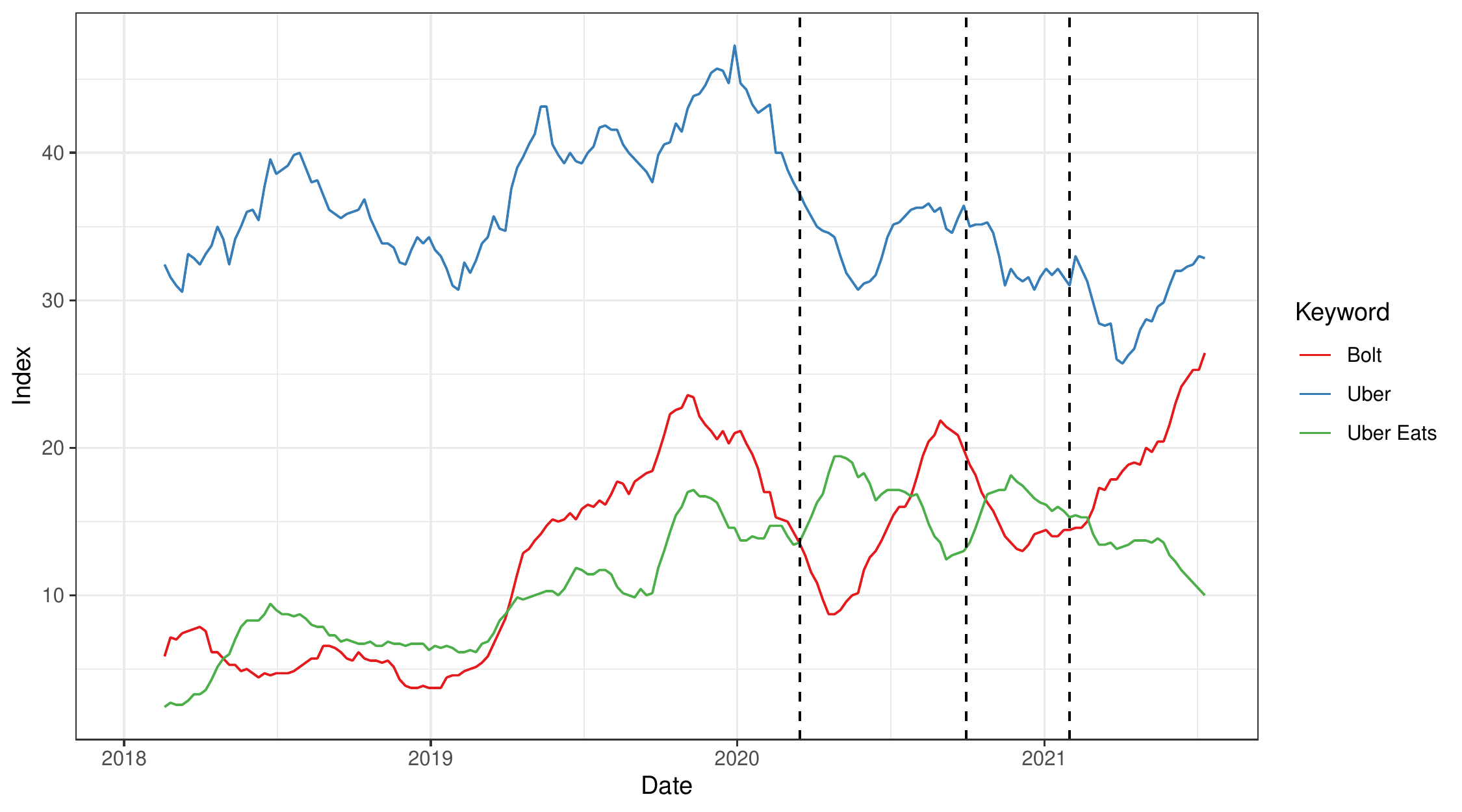}
    \caption{Popularity of two keywords regarding transport and one keyword regarding delivery services in Google Trends. Each line denotes a 7-day moving average and dashed horizontal lines denote the beginning of a given COVID-19 wave. Note that the search word "Bolt" may refer to Bolt's electric scooter or Bolt Food, which was introduced in April 2021}
    \label{gtrends-3}
\end{figure}

\clearpage

\section{Background information about apps users}

\begin{table}[ht!]
\centering
\caption{Background information about the gig economy apps users before (2019HY2) and after (2020HY2) first COVID-19 wave}
\label{tab-demo-users}
\begin{tabular}{llrrrrr}
  \hline
Characteristic & Period & Bolt & Glover & Takeaway & Uber & Wolt \\ 
  \hline
  \multicolumn{7}{c}{Sex} \\ 
  \hline
  Men & 2019 HY2 & 86.9 & 94.4 & 90.2 & 87.9 & 93.6 \\ 
        & 2020 HY2 & 86.3 & 93.8 & 89.7 & 88.0 & 92.3 \\ 
  \cline{2-7} 
  Women & 2019 HY2 & 13.1 & 5.6 & 9.8 & 12.1 & 6.4 \\ 
        & 2020 HY2 & 13.7 & 6.2 & 10.3 & 12.0 & 7.7 \\ 
\hline
  \multicolumn{7}{c}{Age} \\ 
  \hline
  18-30 & 2019 HY2 & 57.0 & 95.0 & 95.5 & 49.1 & 97.8 \\ 
        & 2020 HY2 & 56.2 & 94.1 & 94.6 & 49.2 & 95.7 \\ 
  \cline{2-7}
  31-50 & 2019 HY2 & 37.3 & 5.0 & 3.9 & 46.5 & 2.2 \\ 
        & 2020 HY2 & 37.8 & 5.7 & 4.5 & 46.4 & 2.9 \\ 
  \cline{2-7}
  51-64 & 2019 HY2 & 5.6 & 0.0 & 0.6 & 4.4 & 0.0 \\ 
        & 2020 HY2 & 6.0 & 0.2 & 0.9 & 4.4 & 1.4 \\
  \hline
  \multicolumn{7}{c}{Country of origin} \\ 
  \hline
  Poland & 2019 HY2 & 66.4 & 62.2 & 63.0 & 66.1 & 55.3 \\ 
         & 2020 HY2 & 65.5 & 61.6 & 62.1 & 66.0 & 54.2 \\ 
  \cline{2-7}
  Ukraine & 2019 HY2 & 25.6 & 27.2 & 30.7 & 24.2 & 27.7 \\ 
          & 2020 HY2 & 26.1 & 27.5 & 31.1 & 24.1 & 28.4 \\ 
 \cline{2-7}    
  Other & 2019 HY2 & 8.0 & 10.6 & 6.3 & 9.7 & 17.0 \\ 
        & 2020 HY2 & 8.5 & 10.8 & 6.7 & 9.8 & 17.5 \\ 
  \hline
  \multicolumn{7}{c}{Place of residence} \\ 
  \hline
  Cities & 2019 HY2 & 46.7 & 93.0 & 96.8 & 66.5 & 90.7 \\ 
         & 2020 HY2 & 43.4 & 91.1 & 97.5 & 60.5 & 89.6 \\ 
\cline{2-7}  
  Functional urban areas & 2019 HY2 & 24.4 & 3.2 & 2.6 & 19.5 & 3.5 \\ 
                  & 2020 HY2 & 20.7 & 3.3 & 2.5 & 21.3 & 4.2 \\ 
\cline{2-7}     
  Provinces    & 2019 HY2 & 28.9 & 3.8 & 0.6 & 14.0 & 5.8 \\ 
              & 2020 HY2 & 35.9 & 5.6 & 0.0 & 18.2 & 6.2 \\ 
   \hline
\end{tabular}
\begin{flushleft}
Comment: We only had access to characteristics calculated for a given half-year (HY2) of a~given year and we could not distinguish between units observed in one or both periods.  All variables are the output of the company's proprietary classification or heuristic algorithms, so the level of errors associated with these methods is~unknown. Unfortunately, we did not have access to characteristics of DeliGoo couriers. A~functional urban area is~an area in which at least 15\% of the resident population commutes to work to the city center. For more details, please consult \citet{beresewicz2021}.
\end{flushleft}
\end{table}

\clearpage

\section{Detailed results}\label{supp-results}

\subsection{DeliGoo}

\begin{figure}[ht!]
    \centering
    \includegraphics[width=\textwidth]{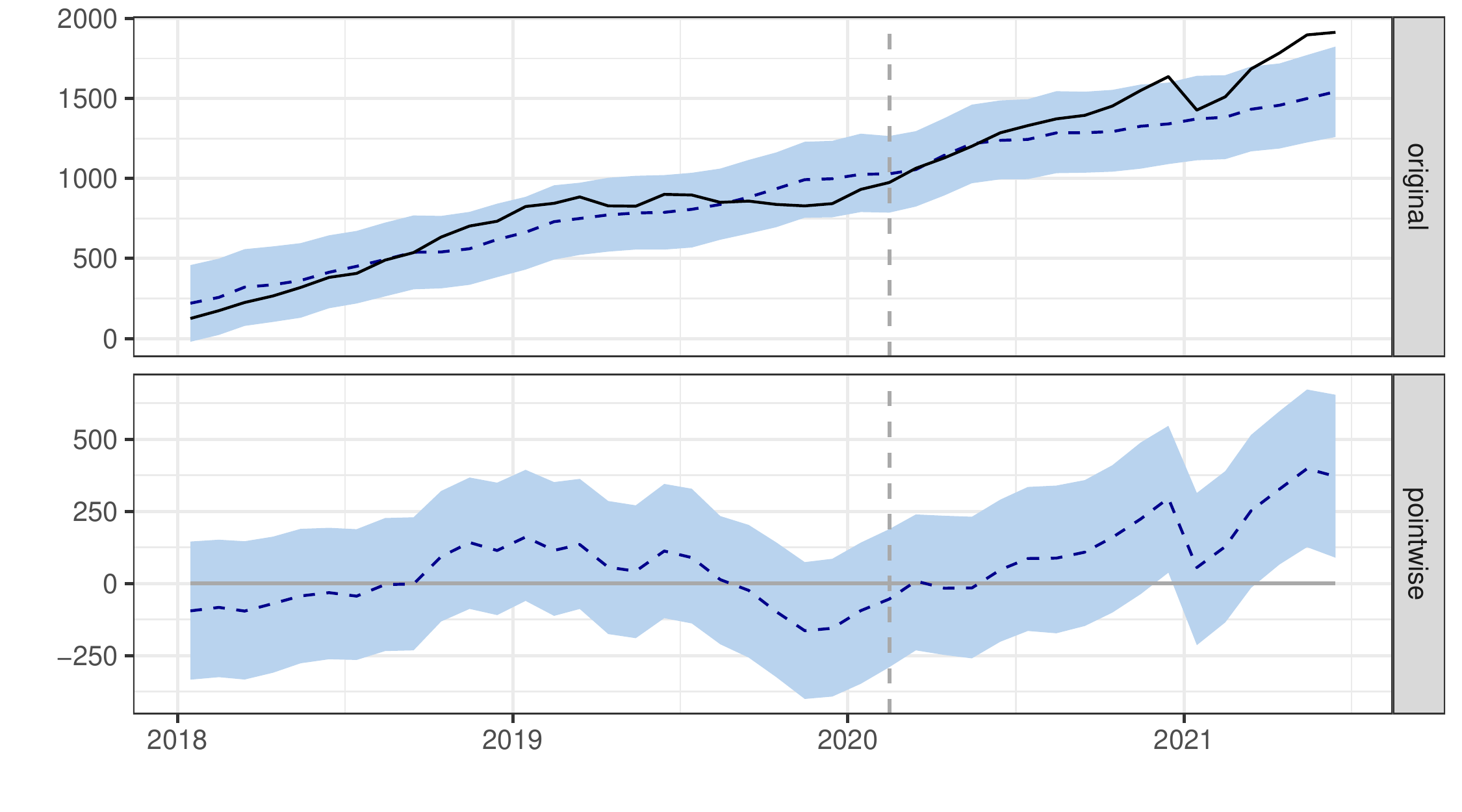}
    \caption{Estimation results for the DeliGoo app based on Bayesian structural time-series models. The blue ribbon denotes a 95\% posterior credible interval}
\end{figure}

\begin{lstlisting}[caption={Output of CausalImpact function for DeliGoo}]
    Posterior inference 
                         Average        Cumulative    
Actual                   1478           23642         
Prediction (s.d.)        1319 (70)      21102 (1124)  
95% CI                   [1184, 1452]   [18939, 23235]

Absolute effect (s.d.)   159 (70)       2540 (1124)   
95% CI                   [25, 294]      [407, 4703]   

Relative effect (s.d.)   12% (5.3%)     12% (5.3%)    
95% CI                   [1.9%, 22%]    [1.9%, 22%]   

Posterior tail-area probability p:   0.01528
Posterior prob. of a causal effect:  98.472%
\end{lstlisting}

\clearpage
\subsection{Glover}

\begin{figure}[ht!]
    \centering
    \includegraphics[width=\textwidth]{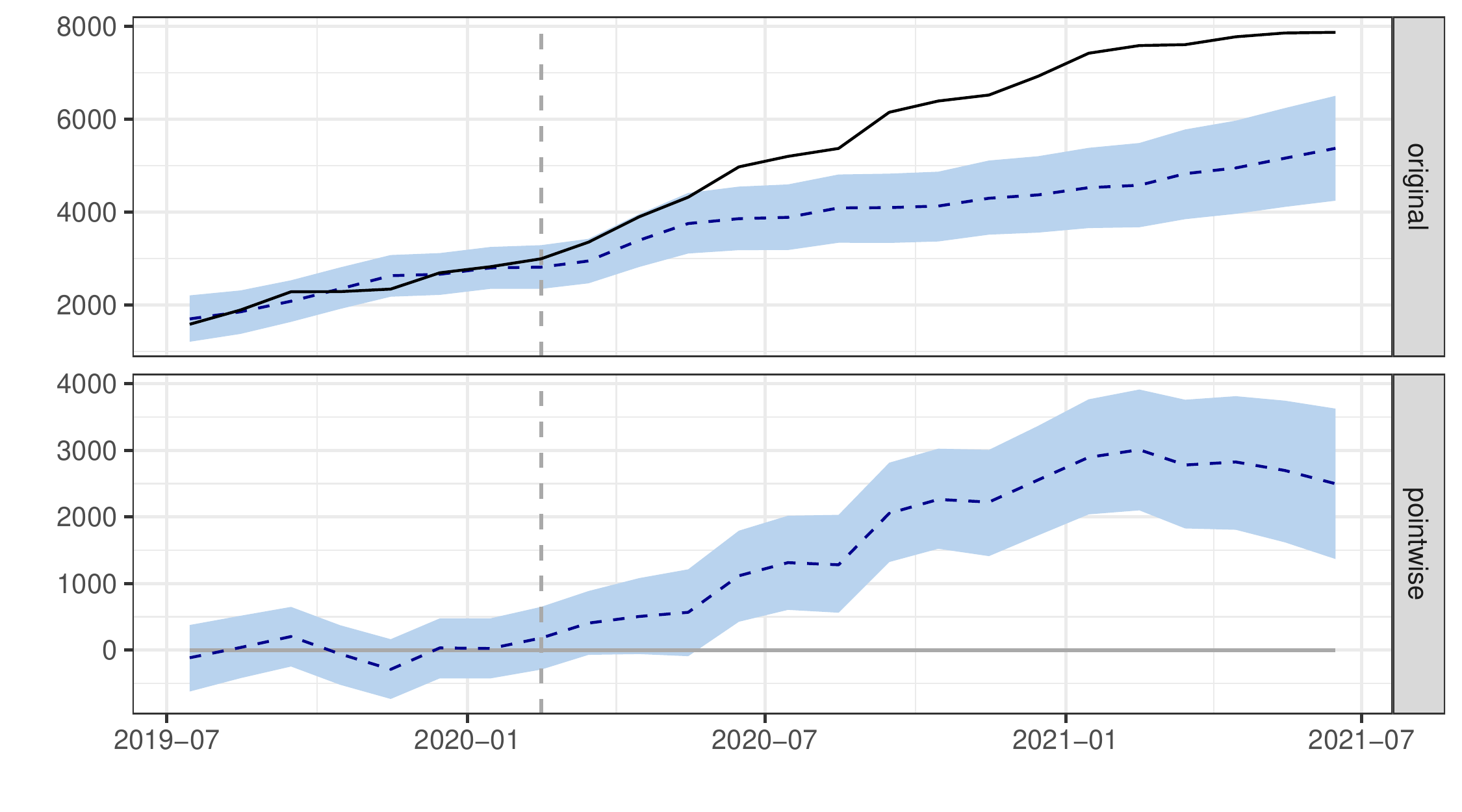}
    \caption{Estimation results for the Glover app based on Bayesian structural time-series models. The blue ribbon denotes a 95\% posterior credible interval}
\end{figure}

\begin{lstlisting}[caption={Output of CausalImpact function for Glover}]
Posterior inference 
                         Average        Cumulative    
Actual                   6202           99237         
Prediction (s.d.)        4257 (346)     68107 (5539)  
95% CI                   [3567, 4958]   [57066, 79334]

Absolute effect (s.d.)   1946 (346)     31130 (5539)  
95% CI                   [1244, 2636]   [19903, 42171]

Relative effect (s.d.)   46% (8.1%)     46% (8.1%)    
95% CI                   [29%, 62%]     [29%, 62%]   

Posterior tail-area probability p:   0.00103
Posterior prob. of a causal effect:  99.89659%
\end{lstlisting}

\clearpage
\subsection{Wolt}

\begin{figure}[ht!]
    \centering
    \includegraphics[width=\textwidth]{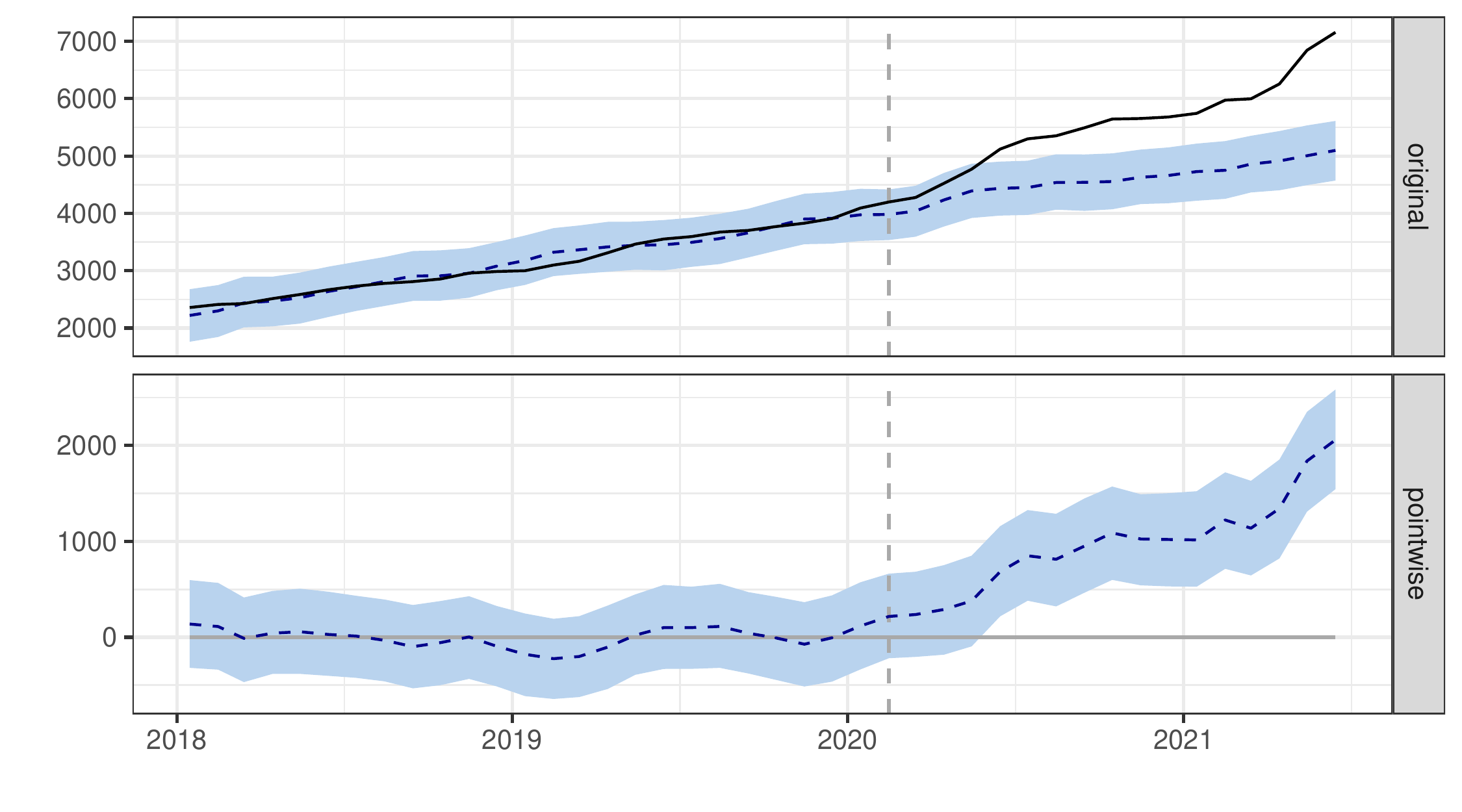}
    \caption{Estimation results for the Wolt app based on Bayesian structural time-series models. The blue ribbon denotes a 95\% posterior credible interval}
\end{figure}

\begin{lstlisting}[caption={Output of CausalImpact function for Wolt}]
Posterior inference 
                         Average        Cumulative    
Actual                   5613           89801         
Prediction (s.d.)        4612 (134)     73788 (2150)  
95% CI                   [4342, 4873]   [69469, 77963]

Absolute effect (s.d.)   1001 (134)     16013 (2150)  
95% CI                   [740, 1271]    [11838, 20332]

Relative effect (s.d.)   22% (2.9%)     22% (2.9%)    
95% CI                   [16%, 28%]     [16%, 28%]    

Posterior tail-area probability p:   0.00103
Posterior prob. of a causal effect:  99.89744%
\end{lstlisting}

\clearpage
\subsection{Bolt}

\begin{figure}[ht!]
    \centering
    \includegraphics[width=\textwidth]{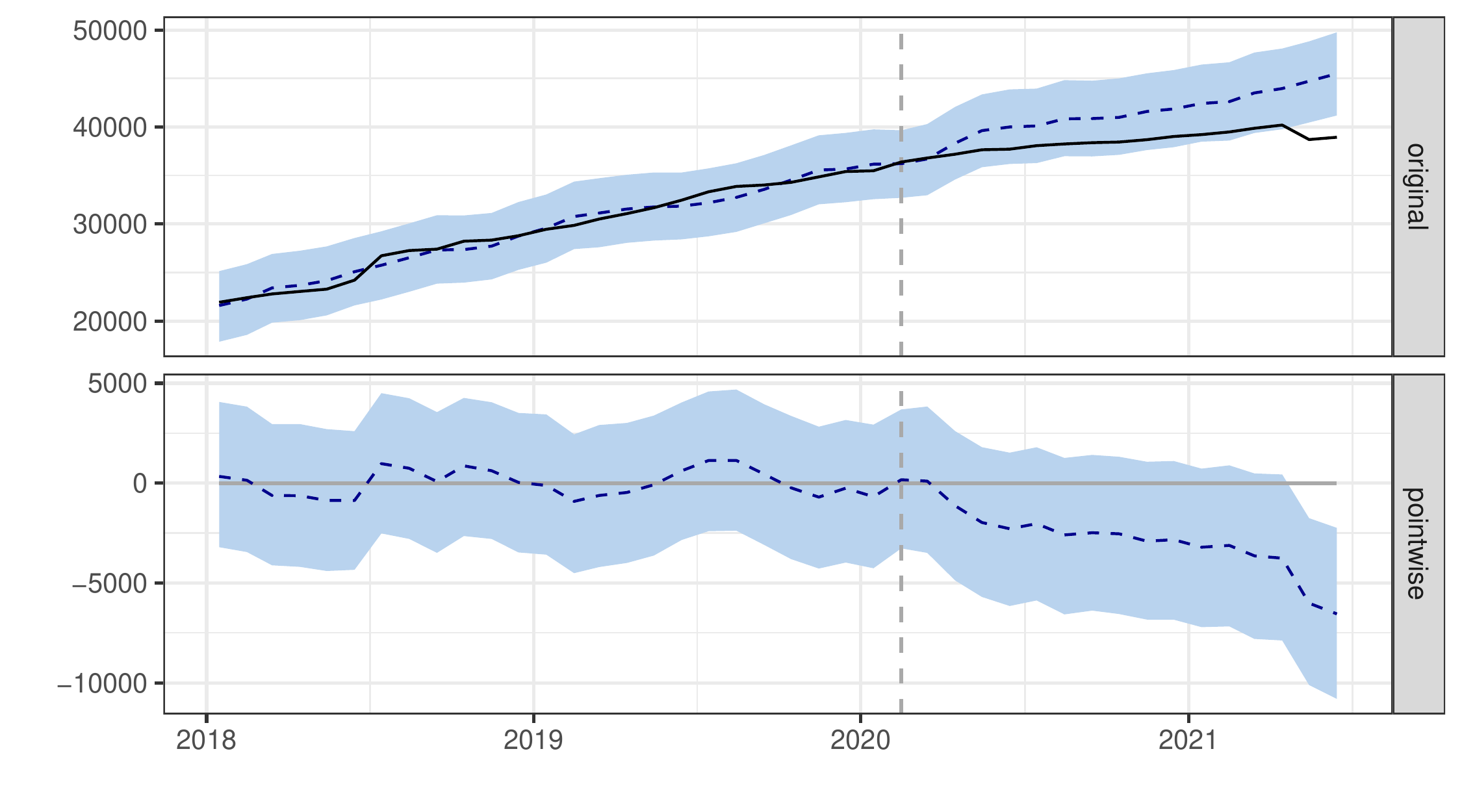}
    \caption{Estimation results for the Bolt app based on Bayesian structural time-series models. The blue ribbon denotes a 95\% posterior credible interval}
\end{figure}

\begin{lstlisting}[caption={Output of CausalImpact function for Bolt}]
Posterior inference 
                         Average        Cumulative    
Actual                   5613           89801         
Prediction (s.d.)        4612 (134)     73788 (2150)  
95% CI                   [4342, 4873]   [69469, 77963]

Absolute effect (s.d.)   1001 (134)     16013 (2150)  
95% CI                   [740, 1271]    [11838, 20332]

Relative effect (s.d.)   22% (2.9%)     22% (2.9%)    
95% CI                   [16%, 28%]     [16%, 28%]    

Posterior tail-area probability p:   0.00103
Posterior prob. of a causal effect:  99.89744%
\end{lstlisting}

\clearpage
\subsection{Uber}

\begin{figure}[ht!]
    \centering
    \includegraphics[width=\textwidth]{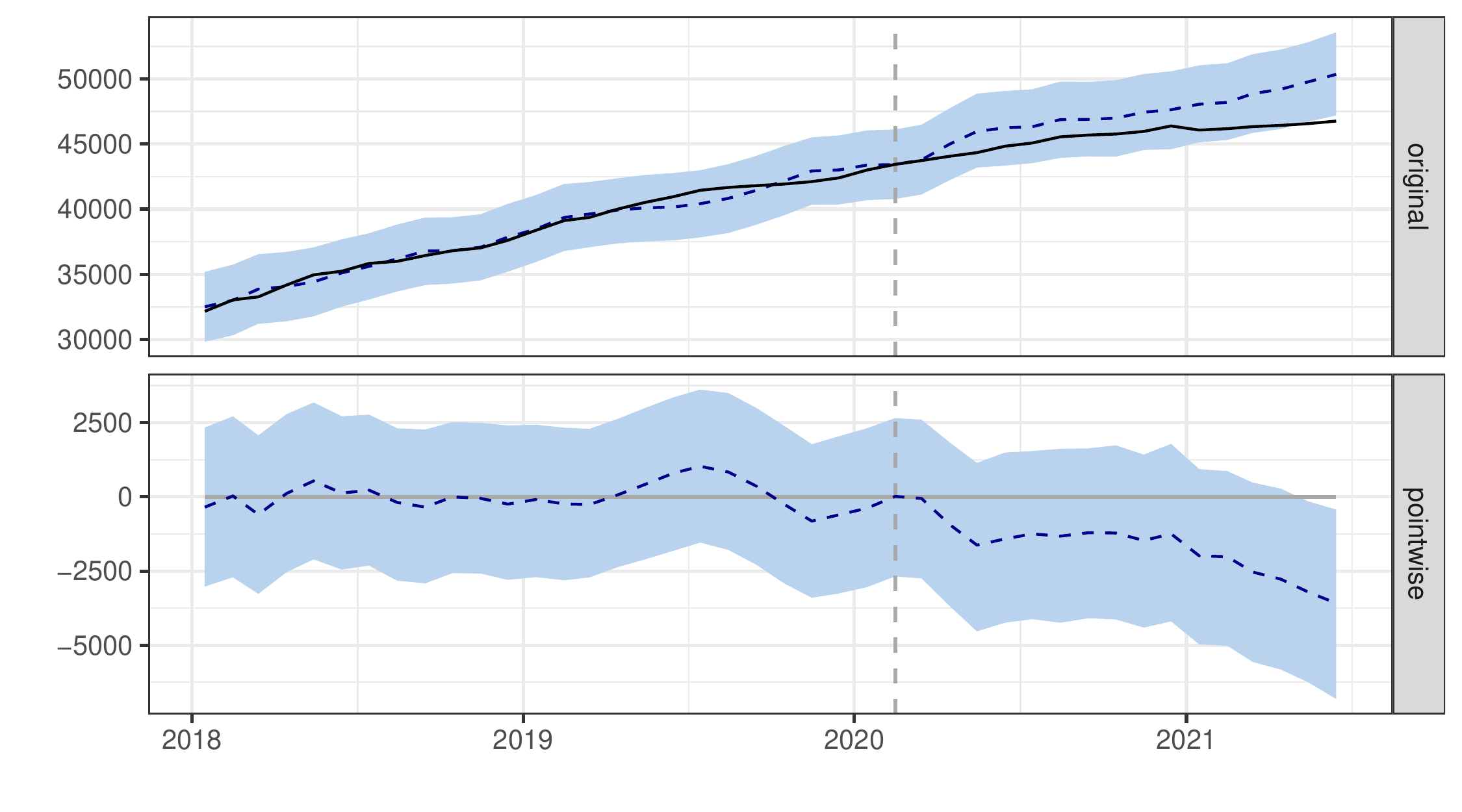}
    \caption{Estimation results for the Uber app based on Bayesian structural time-series models. The blue ribbon denotes a 95\% posterior credible interval}
\end{figure}

\begin{lstlisting}[caption={Output of CausalImpact function for Uber}]
Posterior inference
                         Average          Cumulative      
Actual                   45612            729797          
Prediction (s.d.)        47333 (795)      757329 (12719)  
95% CI                   [45801, 48839]   [732816, 781426]
                                                          
Absolute effect (s.d.)   -1721 (795)      -27532 (12719)  
95% CI                   [-3227, -189]    [-51629, -3019] 
                                                          
Relative effect (s.d.)   -3.6% (1.7%)     -3.6% (1.7%)    
95% CI                   [-6.8%, -0.4%]   [-6.8%, -0.4%]  

Posterior tail-area probability p:   0.01641
Posterior prob. of a causal effect:  98.359%
\end{lstlisting}


\end{document}